\documentclass[12pt]{article}
\usepackage{epsfig}

\begin{document}

{\bf \sf \sc Interstellar scintillation as the origin of
rapid radio variability in the quasar J1819+3845.}\\
\vspace{.1cm}

\centerline{J. Dennett-Thorpe$^{1,3}$ and A.G. de Bruyn$^{2,1}$}

\noindent 1. Kapteyn Astronomical Institute, University of Groningen, 9700 AV, Groningen, The Netherlands\\
2. ASTRON,P.O.Box 2, 7990 AA, Dwingeloo, The Netherlands\\
3. Astronomical Institute Anton Pannekoek, University of Amsterdam, 1098 SJ, Amsterdam, The Netherlands\\

{\bf \sf 
Quasars shine brightly due to the liberation of gravitational energy
as matter falls onto a supermassive black hole in the centre of a
galaxy\cite{ree84}. Variations in the radiation received from active
galactic nuclei (AGN) are studied at all wavelengths, revealing the
tiny dimensions of the region and the processes of fuelling the black
hole.  Some AGN are variable at optical and shorter wavelengths, and
display radio outbursts over years and decades\cite{all85}. These AGN
often also show faster variations at radio wavelengths (intraday
variability, IDV)\cite{hee84,qui89} which have been the subject of
much debate\cite{wag95}. The simplest explanation, supported by a
correlation in some sources between the optical (intrinsic) and faster
radio variations\cite{wag96,pen00}, is that the rapid radio variations
are intrinsic.  However, this explanation implies physically difficult
brightness temperatures, suggesting that the variations may be due to
scattering of the incident radiation in the interstellar medium of our
Galaxy\cite{ric84,bla86}. Here we present results which show
unambiguously that the variations in one extreme case\cite{den00} are
due to interstellar scintillation. We also measure the transverse
velocity of the scattering material, revealing a surprising high
velocity plasma close to the Solar System.}

\vspace{0.2cm}

{ The most extremely variable extragalactic source known at radio
wavelengths is the quasar J1819+3845. This quasar shows variations of
factors of 4 or more on a timescale of hours, easiest interpreted as
scintillation due to scattering in the interstellar medium
(ISM)\cite{den00}. One can think of this scattering medium as focusing
and defocusing the waves from the source, producing a pattern of dark
and bright patches\cite{nar92}. As we move through these patches, or
scintles, we observe a temporal variation in intensity. In order to
display this effect, the source must be small: J1819+3845 is only tens
of microarcseconds (a few light-months at the source redshift
0.54)\cite{den00}. }

\begin{figure}
\epsfxsize=7cm \epsfbox{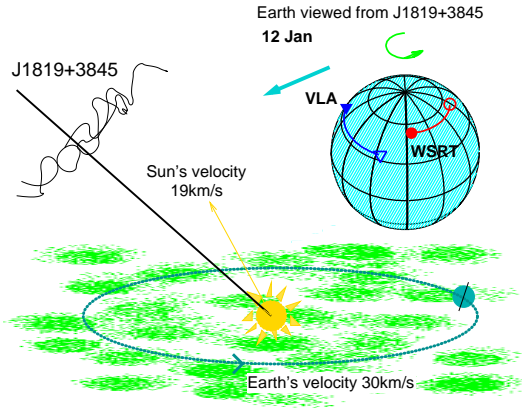}
\caption{ Sketch illustrating the scintles projected onto the solar
system. The inset shows the Earth and its projected velocity
w.r.t. LSR (blue arrow) as seen from J1819+3845. Filled and open
symbols show the telescopes at start and end of the observation run
respectively. As the Earth rotates, the value of the `delay' in
arrival of intensity extrema will change during the observation due to
the changing orientation of the baseline between the telescopes with
respect to the vector velocity of the scintillation pattern. The delay
is also dependent on the time of year (due to the Earth's movement
around the Sun) and the relative motion of the Solar System and the
scattering plasma. }
\label{fig:sketch}
\end{figure}

As the Earth moves through the scintillation pattern, two telescopes
At different locations should see a difference in the observed time of
the intensity maxima and minima (fig.\ref{fig:sketch}). A similar
technique was used by Rickett \& Lang\cite{ric70} who used the
decorrelation of signals between telescopes to put limits on the size
and apparent velocity of the scintles from a pulsar and by Jauncey et
al\cite{jau00} for the other known intra-hour variable
PKS\,0405-387\cite{ked97}.  Compared to other IDV sources, or even
PKS\,0405-387, the rapidity and depth of the variations in J1819+3845
make it the perfect candidate for measurement of this delay.

\begin{figure}
\epsfxsize=13cm \epsfbox{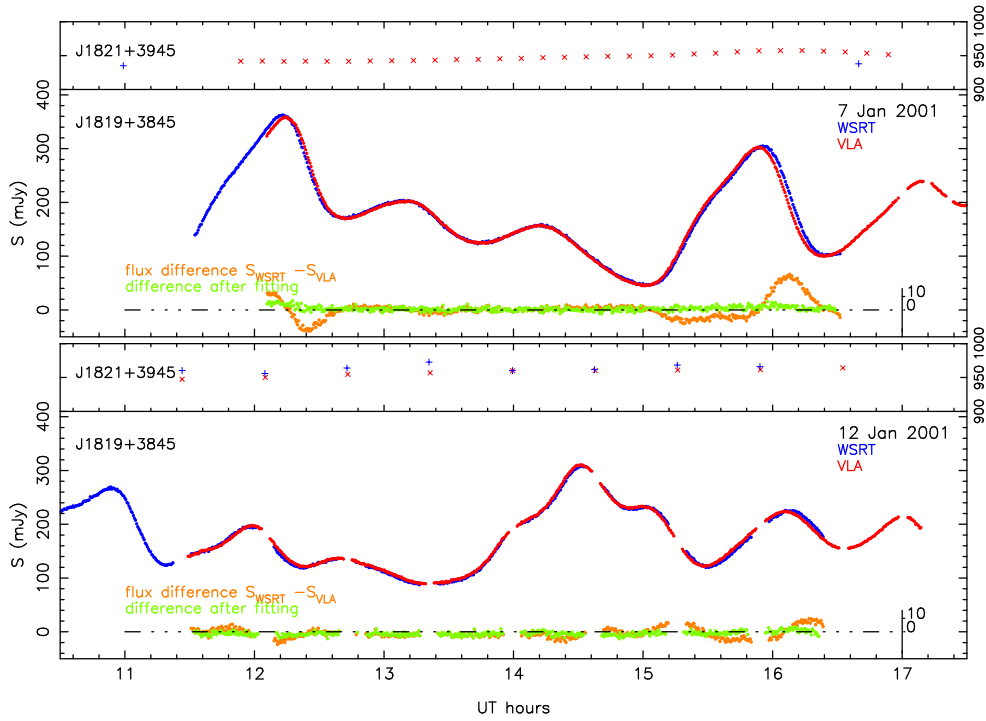}
\caption{Results of VLA and WSRT observations from 7 and 12 January
2001.  Data were averaged over 30\,sec. The time difference in the
arrival of the signals can be seen as a horizontal separation of the
blue and red lines. The vertical axis of the difference signal before
and after fitting has been expanded by a factor of 2.5.  A total of 11
VLA antenna were used for the observations described here.  The
central frequency at both telescopes was 4875\,MHz, and the bandwidths
similar (80 and 100\,MHz). VLA observations of a nearby calibrator
source (J1821+3945) are shown in the top panel of each plot. Slightly
different observing strategies were used on each run. On 7 January
2001 we observed the source continuously, with a sub-set of antennae
observing the calibrator. On 12 January 2001 we used the same antennae
to nod between the source and the calibrator, observing the latter for
3 minutes every $\sim$ 35 minutes. On both dates the observations
described above were bracketed with observations of the calibrator and
a flux density calibrator (3C\,286). Elevation (and other time
dependent) gain variations at the VLA were determined using the
calibrator J1821+3945 were found to be up to 3\%. The data was
corrected for these effects (see below). Observations in good weather
at WSRT typically yield 1\% flux density reliability. This is
applicable to all but the first hour of the 7 January 2001 run, when
light rain caused an estimated extra 1-2\% extinction.  The error we
quote on the result is obtained using a number of different
determinations of the absolute flux density scale, and acceptable
changes throughout the run, including: no temporal changes in gain
corrections, a linear change, and gain corrections determined by
observations of the calibrator, which is assumed stable. Thermal noise
in 30\,sec is 1\,mJy (WSRT) and 0.6\,mJy (VLA). }
\label{fig:results}
\end{figure}

The observations were conducted simultaneously with the NRAO Very
Large Array (VLA) in New Mexico, and the Westerbork Synthesis Radio
Telescope (WSRT) in the Netherlands.  The source showed large, rapid
variations on both days (fig.~\ref{fig:results}). We see that there is
a difference in the time at which the earliest peaks reach the
telescopes (WSRT leads VLA). We also see that at the end of the
observations this delay has reversed, such that the VLA leads
WSRT. This delay, and reversal, is repeated on both days.  If the
scattering plasma is stationary with respect to the Local Standard of
Rest, the predicted time lag from the motion of the Earth and Sun
alone is $\sim$ 200 secs, with the signal arriving at the VLA first
throughout the run. This clearly cannot account for the observations,
so we used the observations to solve for the velocity of the
scattering plasma w.r.t. the LSR, transverse to the line of sight.

\begin{figure}
\epsfxsize=13cm \epsfbox{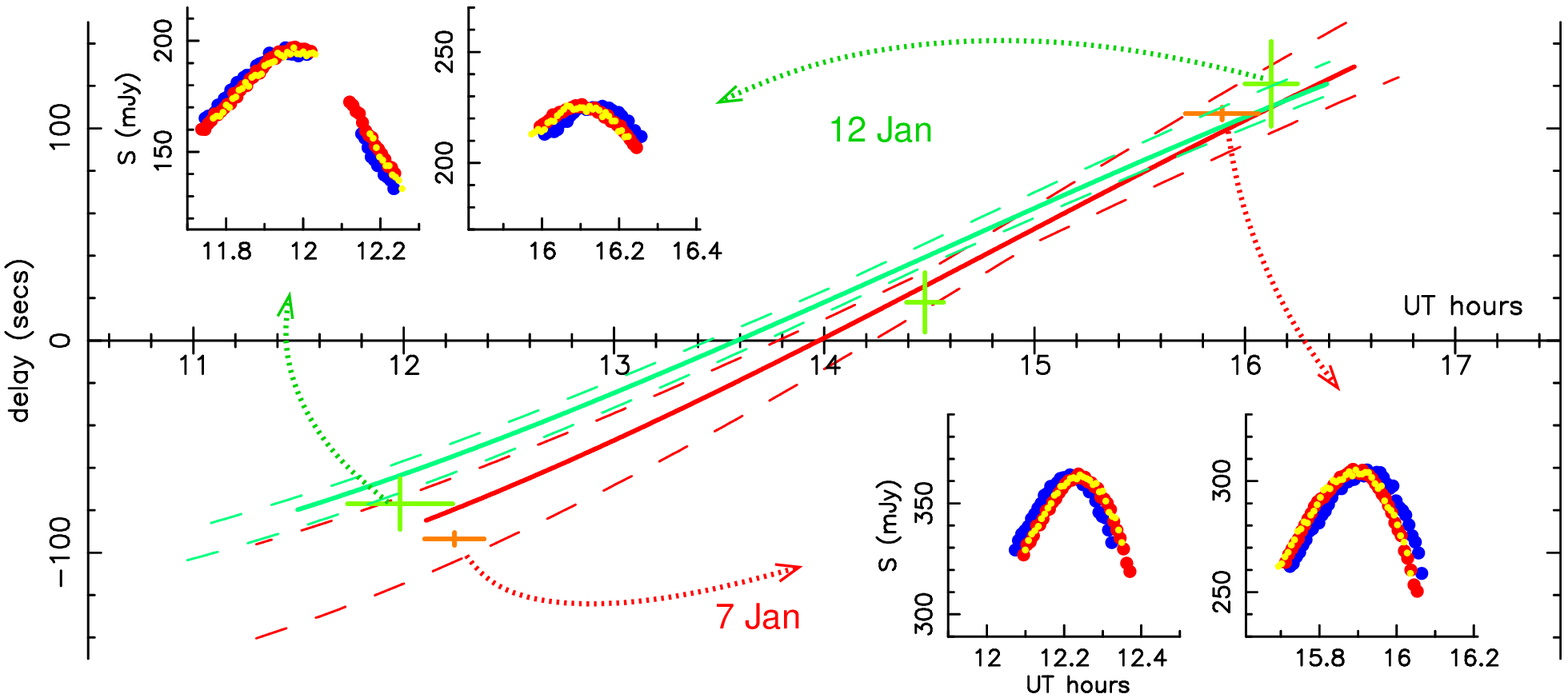}
\caption{\small The calculated delays as a function of observing time for
both days.  The delays are calculated as a function of time from the
projected baseline and velocity vector. We use a circular Earth orbit
and assume that the Sun is moving at 19.7 km/s towards 18h, 30$^\circ$
(J2000). As the direction of the Sun's motion happens to be only
10$^\circ$ from J1819+3845, errors caused by its uncertainty are
negligible. For a wide range of plasma velocities we calculated the
total squared flux density difference, after time shifting the data
according to the calculated delay. Minimizing this quantity gives the
plasma velocity. The associated delays are plotted as the diagonal
lines.
 The results from a given calibration scheme produce errors
of less than 1 km/sec, but the final error is dominated by calibration
uncertainty. The error we quote is that obtained using a number of
different determinations of the absolute flux density scale, and
acceptable changes throughout the run. 
In addition, independent fits to portions of the data are shown with 3$\sigma$
error bars. The data from times with large delays are shown in the
insets (red and dark blue as before, yellow showing the WSRT data
after fitting).}
\label{fig:delay}
\end{figure}

From a minimization of the total squared flux density difference we
calculate the projected plasma velocity.  We obtain for 7 Jan 2001:
v$_{RA} = -32.8 \pm 0.5 $, v$_{Dec} = 14.3 \pm 3 $ km/s and for 12 Jan
2001: v$_{RA} = -32.2 \pm 0.5 $, v$_{Dec} = 16.1 \pm 1$ km/s. It can
be seen that we find very good agreement between the results obtained
from observations of both days. This shows that to first order the
assumption of a temporally 'frozen' scintillation pattern is
justified. Combining these we calculate the plasma velocity to be
v$_{RA} = -32.5 \pm 0.5 $, v$_{Dec} = 15.5 \pm 1 $ km/s (i.e. the
plasma moves West and North).  We also fitted a single time delay to
$\sim$ 15 minutes of data around the largest maxima, normalising the
two lightcurves over the time of interest to the same mean flux
density. Solutions which pass through these values are consistent with
the results quoted above. We plot both these results and the delays
fitted as described above in fig.  ~\ref{fig:delay}.  We can constrain
the plasma velocity so well because at these values the predicted
delay changes very rapidly with changes in plasma velocity. The
existence of many maxima and minima in our data is important as it
removes the possibility of the delay, and the calculated plasma
velocity, being an artefact of flux calibration errors.

We have previously suggested\cite{den00p} a velocity of the plasma
screen in order to explain the change in timescale of the variations
over a year, but this was not a measurement as here, but an
interpretation of the data, and the velocity was very poorly
constrained. For comparison the preliminary result presented in
\cite{den00p}, v$_{RA}$=3 km/s, v$_{Dec}$=25 km/s predicts a very
different delay: from around +60\,sec at the beginning of the
observations to around +150\,sec at the end. We will present a
combination of both these analyses in a forthcoming paper.

We point out that we are measuring the transverse velocity of the
local plasma: in general, radial velocities are measured (via
spectroscopic techniques).  Also note that whereas scintillation
observations of pulsars are significantly affected by their own
transverse velocities (up to several hundred km/s), the effect of a
transverse velocity of (jet features in) AGNs even as high as the
speed of light is reduced to a fraction of a km/sec by the distances
involved.

\vspace{0.2cm}

The determination of a peculiar velocity argues that the scattering
structure is discrete. This measurement of the transverse velocity of
the ionised scattering material cannot be directly compared to radial
velocity measurements of HI, but we note that there is no anomalous
feature known in this direction\cite{har97}. A nearby pulsar
J1813+4013 has an unremarkable dispersion and rotation measures
(42\,cm$^{-3}$pc and 47\,rad\,m$^{-2}$)\cite{ham87}, although the
medium in front of J1819+3845 may not extend to this line of sight.
We have previously estimated\cite{den00} that this scattering region
is located at around 20\,pc from the Sun. The scatterer may be related
to the Local Bubble\cite{bha98}, and if so, for the reasons given
above, is more likely to be a discrete structure at the boundary than
due to the putative intervening hot gas.

A small difference in the flux densities observed at the two
telescopes (after delay correction) as the telescopes pass through
different regions of the scintles is expected. The size of the
scintles in the direction of motion is $\sim$ 4000\,sec $\times$ 25
km/sec = 10$^5$km. If the scintles were of similar size perpendicular
to this, we can estimate the expected difference in the intensities on
a typical baseline of 6000\,km by considering the flux density
differences between the two curves if they are shifted
6000/10$^{5}\times 4000 = 240$sec. This gives a potential 5-10\% flux
density difference (after delay correction) between the two
curves. However, differences are less than 2\% on both days
(fig.\ref{fig:results}), suggesting the scintillation pattern is
highly anisotropic, in agreement with preliminary analysis of the
variation of the timescale over a year\cite{den00p}.

To account for these observations there must be an unusually high
fraction of turbulent electrons, if the electron density is that of
the typical extended ISM. In comparison to the conditions found in
highly turbulent, ionised regions, e.g. around OB
stars\cite{tc93,cor85,mor90} however, the required conditions are very
modest.  Interestingly, the A0 star $\alpha$ Lyrae (Vega), which at
8\,pc is within the estimated distance range for the screen, is at a
projected angular separation of J1819+3845 of 3$^\circ$, which is only
0.4\,pc.  Whether the origin of the turbulent plasma could be related
to mass loss from this star, or a feature of the so-called Local
Fluff\cite{cox87}, remains to be investigated.

\vspace{0.2cm}

Our observations prove beyond doubt that the variations in J1819+3845,
the most extreme example of an IDV source, are due to scintillation.
Yet, as this source must be extremely small, it could easily be
intrinsically variable on time scales of a few months (using simple
light-travel time arguments).  Using relativistic outflow arguments
its is not difficult to bring this down to weeks.  Intrinsic
variations on these time scales would be very hard to separate from
scintillation induced variations.  We have evidence however that the
source is remarkably stable over a period of many months with only a
slow increase in brightness over the last two years (Dennett-Thorpe \&
de Bruyn, in prep). Scintillation in known variable sources and those
with slower variations will be more difficult to prove directly.
However, the few percent variations seen on day scales in many
IDV\cite{qui92} is likely to be due to scattering of structures of
tens of microarcseconds through more typical ISM\cite{ric95}.  This
implies that most AGN have brightness temperatures no higher than
$10^{12}-10^{14}$K, not $10^{17} - 10^{20}$K, as would otherwise be
required.  Perhaps the most serious argument against an intrinsic
explanation of the IDV phenomenon, which requires extremely small
angular sizes, now becomes how to avoid scintillation induced
variations dominating all the observed variability.

The lack of correlation of strength or timescale of variations with
frequency or galactic latitude is sometimes cited as evidence against
ISS as cause for IDV\cite{kra99}. However as the variations are
affected by the intrinsic size of the source, the location and motion
of the effective scatterer as well as the season of the observations,
such correlations may easily be washed out.  Although the annual
modulation in the variations of B\,0917+62 have been explained in
terms of scintillation through the general Galactic
ISM\cite{ric01,jau01}, J1819+3845 demonstrates that one cannot assume
that the scattering effectively occurs throughout this medium. It is
important to note that the scattering of resolved AGN will be weighted
by the material closer to the observer.

The delay in arrival times of the intensity variations at two widely
separated radio telescopes is a neat proof that a propagation effect
causes the extreme variations in this one quasar. The small size of
the quasar J1819+3845 is probably related to the unusual rising radio
spectrum.  The absence of many sources which scintillate in this
manner is presumably due to a combination of an unusually small,
stable source, and an atypically strong, close scatterer. We
nonetheless expect that more similarly extreme IDVs will be
discovered.

\noindent{\bf Acknowledgements}

We are grateful to Frank Briggs for the code used to predict the
delays. This work was funded by the EU TMR network `CERES'. The
National Radio Astronomy Observatory is a facility of the National
Science Foundation operated under cooperative agreement by Associated
Universities. The WSRT is operated by ASTRON, with financial support
by the Netherlands Organization for Scientific Research (NWO). 

\vspace {0.2cm}

\noindent Correspondence and requests for materials should be addressed to
ger@astron.nl.

\end{document}